\documentclass[12pt]{article}
\usepackage{amssymb}
\usepackage{graphicx}
\usepackage{epsf}
\usepackage{bm}

\newcommand{\gsm}{~g~cm$^{-3}$\,}

\begin{document}

\title {Shear Viscosity and Oscillations of Neutron Star Crusts}
\author{A.I. Chugunov and D.G. Yakovlev\\
 {\it\small Ioffe Physico-Technical Institute, St. Petersburg, Russia}}
\date{}
 \maketitle

\begin{abstract}
We calculate the electron shear viscosity
(determined by Coulomb electron collisions)
for a dense matter in a wide range of
parameters typical for white dwarf cores and neutron star crusts.
In the density range from $\sim 10^3$ g~cm$^{-3}$ to
$10^7-10^{10}$ g~cm$^{-3}$ we consider the matter composed of
widely abundant astrophysical elements, from H to Fe. For higher
densities, $10^{10}-10^{14}$ g~cm$^{-3}$, we employ the
ground-state nuclear composition, taking into account finite sizes
of atomic nuclei and the distribution of proton charge over the
nucleus. Numerical values of the viscosity are approximated by an
analytic expression convenient for applications.
Using the approximation of plane-parallel layer we study
eigenfrequencies, eigenmodes and viscous damping times of oscillations
of high multipolarity, $l \sim 500-1000$, localized in the outer
crust of a neutron star. For instance, at $l \sim 500$
oscillations have frequencies $f \gtrsim 40$ kHz and are localized
not deeper than $\sim$300 m from the surface. When the crust
temperature decreases from $10^9$~K to $10^7$~K, the dissipation
time of these oscillations (with a few radial nodes) decreases
from $\sim 1$ year to 10--15 days.
\end{abstract}
\section{Introduction}
\label{SecIntrod}

The shear viscosity of dense stellar matter (at densities
$\rho\lesssim 10^{14}$ g~cm$^{-3}$) is important for a number of
astrophysical problems, including the viscous
damping of oscillations in white dwarfs and in envelopes (crusts) of
neutron stars. The total shear viscosity can be presented as a sum
of various matter components. In the outer crust of a
neutron star or in the core of a white dwarf, it is determined by
electrons and ions, $\eta=\eta_{\rm e}+\eta_{\rm i}$.
In the inner crust of a neutron star
it is
necessary to add a contribution of free neutrons, $\eta_n$.
Under formulated conditions, the electrons are strongly
degenerate and form an ideal Fermi gas, while the ions are fully
or partially ionized and form a strongly nonideal Coulomb fluid or
Coulomb crystal. In this case the electrons become the
most important carriers of heat, charge (see, e.g.,
Ref.\ \cite{Potekhin1999}), and momentum. The main process
of electron scattering
which determines kinetic coefficients (thermal conductivity, electrical
conductivity, and viscosity) is the scattering of electrons by
ions (atomic nuclei).

The shear viscosity of dense stellar matter
determined by electron-ion scattering has been considered in a
number of papers. For example, the electron viscosity of a strongly
nonideal Coulomb liquid was calculated in Refs.\
\cite{Flowers&Itoh1976,Flowers&Itoh1979,Nandkumar&Pethick1984} from
variational principle. The results of these calculations are in good
agreement. However, they were carried out neglecting the
quasi-ordering of ions, which is important near the melting point.
The inclusion of this quasi-ordering in a liquid together with
multiple-phonon process of electron scattering in a crystal led to
the disappearance of appreciable (by a factor of two to four) jumps
in the electrical and thermal conductivities \cite{Potekhin1999}.

Previous calculations of the viscosity were carried out in the Born
approximation. However, the non-Born corrections are important for
calculating the electrical and thermal conductivities of the matter
containing chemical elements with high charge numbers $Z$ (see, e.g.,
Ref.\ \cite{Potekhin1999}). We include these corrections and show
that they are equally important for calculations of the viscosity.

While studying oscillations of neutron star envelopes, it
is necessary to know the viscosity of the matter with the density
$\rho \lesssim 10^{14}$ g~cm$^{-3}$. For $\rho\gtrsim10^{13}$
g~cm$^{-3}$, the sizes of atomic nuclei become comparable to the
distances between them, and it is necessary to take into account
the distribution of proton charge within the nuclei. This effect
was included in the calculation of the electrical and thermal conductivity
\cite{Itoh&EtAll1984, Itoh&EtAll1993} by
introducing form factors of atomic nuclei. Such
calculations have not been carried out yet for the viscosity.

In this paper, we have calculated the shear
viscosity taking into account non-Born corrections and the form
factor of the nuclei, the quasi-ordering in a Coulomb liquid, and
multi-phonon processes in a Coulomb crystal. The results are
approximated by analytic formulas convenient for applications.

Various types of oscillation modes can be excited in neutron stars.
Generally speaking, these oscillations carry important information
on the internal structure of neutron stars. Specific types of
oscillations (such as r modes) can be accompanied by the radiation
of gravitational waves.
%Interest in studies of neutron star
%oscillations has been continuously growing.
Since neutron stars
are relativistic objects, their
oscillations should be analyzed in the framework of General
Relativity. The relativistic theory of oscillations was developed
in a series of papers by Thorne and coauthors
\cite{ThornPaperI,ThornPaperII,ThornPaperIII,ThornPaperIV,ThornPaperV,ThornPaperVI}.
In particular, the rapid ($\sim 1$ s) damping of p-modes
with multipolarity $l = 2$ owing to gravitational
radiation was demonstrated in Ref.\ \cite{ThornPaperIII}. An exact
inclusion of general-relativistic effects is complicated, but, in
many cases, it is possible to use the relativistic Cowling
approximation \cite{McDermott1983}. A comprehensive analysis of various
oscillation modes and mechanisms of their dissipation was carried
out in Ref.\ \cite{McDermott1988}. We also note the recent review of
Stergioulas \cite{sterg03}, which contains an extensive
bibliography. As a rule, oscillations with low values of $l$ have
been considered in the literature.

Although neutron stars are in the final stage of stellar evolution,
they can be seismically active for many reasons. Possible mechanisms
for the generation of oscillations have been widely discussed in the
literature (see, e.g.,
\cite{McDermott1988,sterg03} and references therein). Much attention
has recently been paid to r modes --- vortex oscillations that can
be generated in rapidly rotating neutron stars and are accompanied
by powerful gravitational radiation. In addition, oscillations can
be excited in neutron stars, for example, during X-ray bursts
(nuclear explosions on the surfaces of accreting neutron stars),
bursting activity of magnetars (anomalous X-ray pulsars and soft
gamma-ray repeaters; see, e.g., Ref.\ \cite{kaspi04}), and glitches
(sudden changes of spin periods) of ordinary pulsars.

In this paper, we study the damping of oscillations in order to
illustrate the viscosity calculations. We choose the simplest
example --- p mode oscillations localized in the outer crust due to
a high multipolarity, $l\gtrsim 500$.

\section{Shear viscosity of dense stellar matter}

\subsection{Parameters of dense stellar matter}

Let us consider dense stellar matter which consists of ions (atomic
nuclei) and degenerate electrons. The state of strongly degenerate
electrons can be characterized by their Fermi momentum $p_{\rm F}$
or wave number $k_{\rm F}$,
\begin{equation}
    p_{\rm F}\equiv\hbar k_{\rm F}
    =\hbar\,\left(3\pi^2\,n_{\rm e}\right)^{1/3}
    =m_{\rm e}\, c\, x_{\rm r},
\end{equation}
where $\hbar$ is the Planck constant, $m_{\rm e}$ and $n_{\rm e}$
are the mass and number density of electrons, $x_{\rm r}\approx
1.009\left(\rho_6Z/A\right)^{1/3}$ is the relativistic parameter
of the electrons, $Z$ and $A$ are the charge number and atomic number of
the ions, and $\rho_6$ is the density in units of
$10^6$~g~cm$^{-3}$. The electron degeneracy temperature is
\begin{equation}
    T_{\rm F}=\left(\varepsilon_{\rm F}
    -m_{\rm e}c^2\right)/k_{\rm B}
    \approx 5.93\cdot 10^9\left(\sqrt{1+x_{\rm r}^2}-1\right)~~\mbox{K},
\end{equation}
where $k_{\rm B}$ is the Boltzmann constant and
\begin{equation}
    \varepsilon_{\rm F}
    \equiv m_{\rm e}^\ast c^2
    =m_{\rm e} c^2\sqrt{1+x_{\rm r}^2}
    %\approx 8.187\cdot 10^{-7}\sqrt{1+x_{\rm r}^2}\quad\mbox{ergs}
\end{equation}
is the electron Fermi energy. In our study, we consider the matter at
$T\ll T_{\rm F}$ and $T\lesssim 5\cdot 10^9$~K (the latter is
required to avoid dissociation of atomic nuclei).

Let us also introduce the Fermi velocity of the electrons,
\begin{equation}
    v_{\rm F}\equiv c\beta_{\rm r}= p_{\rm F}/m_{\rm e}^\ast .
\end{equation}
The electrostatic screening of a test charge by degenerate
electrons is described by the Thomas-Fermi wave-number $k_{\rm TF}$
(the inverse electron screening length),
\begin{equation}
    k_{\rm TF}^2=4\pi e^2\,\frac{\partial n_{\rm e}}{\partial \mu} %
        \approx \frac{\alpha}{\pi \beta_{\rm r}} %
            \left(2 k_{\rm F}\right )^2,
\end{equation}
where $\mu\approx\varepsilon_{\rm F}$ is the electron chemical
potential and $\alpha=e^2/\hbar c\approx1/137.036$ is the
fine-structure constant.

The state of ions is characterized by the Coulomb
coupling parameter
\begin{equation}
    \Gamma=\frac{Z^2e^2}{ak_{\rm B}T}\approx\frac{22.75 Z^2}{T_6}
    \left(\frac{\rho_6}{A}\right)^{1/3},
\end{equation}
where $a=\left[ 3/\left(4\pi n_{\rm i}\right)\right]^{1/3}$ is the
ion sphere radius; $n_{\rm i}=n_{\rm e}/Z$ is the number
density of ions; and $T_6$ is the temperature in units of $10^6$~K.
When $\Gamma\ll 1$, the ions form a nearly ideal Boltzmann gas. If
$\Gamma\gtrsim 1$, they form a strongly nonideal Coulomb fluid.
Finally, when $\Gamma>\Gamma_{\rm m}$ (which is realized at
$T < T_{\rm m}$), the ions crystallize. The crystallization of a
classical system of ions occurs at $\Gamma_{\rm m}\approx 175$
(see, e.g., \cite{dewitt01}).

Quantum effects in ion motion become important at
$\Theta\equiv T/T_{\rm p}\ll 1$,  where
\begin{equation}
    T_{\rm p}=\hbar\omega_{\rm p}/k_{\rm B}\approx
    7.832\cdot 10^6(Z/A)\rho_6^{1/2}\mbox{ K}
\end{equation}
is the ion plasma temperature, $\omega_{\rm p}=\left(4\pi
Z^2e^2n_{\rm i}/m_{\rm i}\right)^{1/2}$ is the ion plasma
frequency, $m_{\rm i} = A m_{\rm u}$ is the ion mass, and
$m_{\rm u} = 1.6605 \times 10^{-24}$~g is the atomic mass unit.

For isotropic matter, the viscous stress tensor has the
simple form
\begin{equation}
\label{Eq1}
    \sigma^\prime_{\alpha\beta}
    =\eta\left({\frac{\partial {U_\alpha}}{\partial {x_\beta}}}+{\frac{\partial {U_\beta}}{\partial {x_\alpha}}}
    -\frac 23\, \delta_{\alpha\beta}\,\nabla\cdot {\bm U}\right)
    +\zeta\,\delta_{\alpha\beta}\,\nabla\cdot {\bm U},
\end{equation}
where $\bm U$ is the hydrodynamical velocity of the matter, $\eta$
is the shear viscosity, and $\zeta$ is the bulk viscosity (this last
quantity is especially important for uniform compressions and
rarefactions of the matter).

Generally, crystalline matter is anisotropic, and
Eq.~(\ref{Eq1}) for the viscous stress tensor may
be formally invalid. However, in a dense matter, ions crystallize
into highly symmetric body-centered cubic
(bcc) or face-centered cubic (fcc)
lattice. In this case, the viscous stress tensor for a monocrystal
is determined by three independent coefficients (see, e.g.,
Ref.\ \cite{UprTheory}), and can be written in the form (\ref{Eq1}) with
an additional term of the form $\kappa\,
\delta_{\alpha\beta}\partial{U_\alpha}/\partial{x_\alpha}$ (the
sum over $\alpha$ is not carried out). The quantity $\bm U$ should
be treated as the velocity field for shifts of the ions in their
lattice sites. When studying transport processes on scales
exceeding the characteristic monocrystal size, the matter can be
considered as isotropic. As in all the literature concerned with
the kinetics of crystalline matter of white dwarfs and
neutron stars without magnetic fields, we will restrict our
analysis to this case (assuming  $\kappa=0$).

The shear viscosity of neutron star crusts and
white dwarf cores is primarily determined by strongly degenerate
electrons. It is convenient to present this viscosity in the form
\begin{equation}
    \eta_{\rm e}=\frac{n_{\rm e} p_{\rm F} v_{\rm F}}{5\nu_{\rm e}},
\end{equation}
where $\nu_{\rm e}=1/\tau_{\rm e}$ is the effective electron
collision frequency, which is the inverse of the effective electron
relaxation time $\tau_{\rm e}$. If the electron scattering is
determined by several independent processes, these can be studied
separately, and the total collision frequency will be the sum of
partial ones. For the dense matter of white dwarf cores and neutron
stars envelopes, one has
\begin{equation}
    \nu_{\rm e}=\nu_{\rm ei}+\nu_{\rm imp}     +\nu_{\rm ee},
\end{equation}
where $\nu_{\rm ei}$, $\nu_{\rm imp}$ and $\nu_{\rm ee}$ correspond
to electron scattering by ions, impurity ions, and electrons,
respectively. The dominant process is electron-ion scattering; it will
be studied below. This scattering
determines also the thermal and electrical conductivities of dense matter
(see, e.g., Ref.\ \cite{Potekhin1999}). With small variations, the
formalism proposed by Potekhin {\it et al.} \cite{Potekhin1999} is
applicable for calculations of the viscosity.

In crystalline matter, the electron-ion interaction can adequately
be described in terms of the emission and absorption of phonons
\cite{Ziman1962}. This description can be realized using an ion
dynamical structure factor \cite{Flowers&Itoh1976}.

The frequency of electron-ion collisions ($ei$ scattering) can be
written as
\begin{equation}
\label{Eq2}
    \nu_{\rm ei}=12\pi \frac{Z^2e^2\Lambda_{\rm ei}n_{\rm i}}{p_{\rm F}^2v_{\rm F}}
            =\frac{4Z\varepsilon_{\rm F}}{\pi\hbar}\alpha^2\Lambda_{\rm ei},
\end{equation}
where $\Lambda_{\rm ei}$ is the effective Coulomb logarithm, which
can be calculated using the variational method (see, e.g.,
Ref.\ \cite{Ziman1962}). Using the simplest trial function in the Born
approximation for a strongly nonideal ion plasma ($\Gamma\gtrsim
1$), one obtains
\begin{equation}
\label{Eq3}
    \Lambda_{\rm ei}=\int_{q_0}^{2k_{\rm F}} q^3u^2(q)
    \left(1-\frac{q^2}{4k_{\rm F}^2}\right)
    \left[1-\frac 14 \left(\frac{\hbar q}{m_{\rm e}^\ast
    c}\right)^2\right] S_\eta(q)\,{\rm d}q,
\end{equation}
where $q_0$ is the minimum momentum $q$ transferred in an $ei$
scattering event; $q_0 = 0$ for the liquid phase and $q_0 = q_{\rm
B}$ in the crystalline phase, where $q_{\rm B} = \left(6\pi n_{\rm
i}\right)^{1/3}$ is the radius of a sphere of the same volume as the
Brillouin zone. The value $q_0 = q_{\rm B}$ selects
Umklapp processes (i.e., those involving electron
momentum transfer $q \gtrsim q_{\rm B}$) in an $ei$ scattering event. At not
very low temperatures
\begin{equation}
    T\gtrsim T_{\rm u}\sim T_{\rm p} Z^{1/3}\alpha_f/(3\beta_{\rm r}),
\end{equation}
the contribution of such processes to the
Coulomb logarithm is much higher than the contribution of normal
processes which occur at $q<q_{\rm B}$ (see, e.g., Ref.\
\cite{Baiko&Yakovlev1995}). However, at low temperatures ($T\lesssim
T_{\rm u}$), Umklapp processes are ''frozen out'' and the viscosity is
determined by normal processes. We will neglect this effect below,
restricting our consideration to temperatures $T\gtrsim T_{\rm u}$.

The function $u(q)$ in Eq.~(\ref{Eq3}) describes the Coulomb
interaction between an electron and an atomic nucleus; it is
discussed in Section 2.3. The factor in the square brackets describes
the kinematic effect of backward scattering of
relativistic electrons (see, e.g.,
\cite{Berestetskii2001}); $S_\eta(q)$ is an effective ion static
structure factor that takes into account ion correlations. This
factor coincides with the effective structure factor
that determines
the electrical resistivity of the dense matter (it was
calculated and approximated in Ref.~\cite{Baiko1998}). Note that the
structure factor of a strongly nonideal Coulomb liquid is known
only in the classical limit ($\Theta\gg 1$). We also define a {\it
simplified} structure factor, based on the two
approximations:
(1) Neglecting quasi-ordering in ion positions in the
    Coulomb fluid (see, e.g., Ref.\ \cite{Potekhin1999}).
(2) Single-phonon approximation for the inelastic
    structure factor of the Coulomb crystal
    (see, e.g., Ref.\ \cite{Baiko&Yakovlev1995}).
We will call the viscosity calculated using the simplified structure
factor the {\it simplified} viscosity. Note that previous
calculations of the shear viscosity by Flowers and Itoh
\cite{Flowers&Itoh1976,Flowers&Itoh1979} and by Nandkumar and Pethick
\cite{Nandkumar&Pethick1984} were carried out for a Coulomb liquid
using the simplified structure factor.

To account for non-Born corrections, we also
multiply the integrand by the ratio of the exact and Born cross
sections for Coulomb scattering. This method was proposed in
Ref.\ \cite{Yakovlev1987} and used for calculating the transport
coefficients by Potekhin {\it et al.}
\cite{Potekhin1997,Potekhin1999}.

The effective frequency of electron scattering by impurities
(assuming that impurity atoms are randomly distributed over
the crystal) is similar to the frequency of electron-ion scattering
(see Eq.~(\ref{Eq2})):
\begin{equation}
    \nu_{\rm imp}=\frac{12\pi e^4}{p_{\rm F}^2v_{\rm F}}
        \sum_{\rm imp}\left(Z-Z_{\rm imp}\right)^2
    n_{\rm imp}\,\Lambda_{\rm imp},
\end{equation}
where $Z_{\rm imp}$ is the charge number of an impurity ion and the
Coulomb logarithm $\Lambda_{\rm imp}$ is calculated using
Eq.~(\ref{Eq3}), but assuming that impurities are weakly correlated
(which corresponds to the structure factor $S_{\rm imp}\equiv 1$, while
the screening of impurities is included into the factor
$u(q)$). In the simplest model of Debye screening (with a
screening length $q_{\rm Simp}^{-1}$)
\begin{equation}
    \Lambda_{\rm imp}=\frac12
        \left(1+3\beta_{\rm r}^2\xi^2+2\xi+2\xi\beta_{\rm r}^2\right)
        \ln\left(\frac{1+\xi}{\xi}\right)
        -\frac 32 \,\beta_{\rm r}^2\,\xi-\frac 14\, \beta_{\rm r}^2-1,
\end{equation}
where $\xi=q_{\rm S\,imp}/(2k_{\rm F})$; $q_{\rm S\,imp}^2=k_{\rm
TF}^2+k_{\rm imp}^2$, and $k_{\rm imp}$ is the wave number for the
Debye screening of a test charge by impurities (the inverse
correlation length of impurities). This wave number weakly affects the
Coulomb logarithm ($k_{\rm TF}\gg k_{\rm imp}$), and can be estimated as
$k_{\rm imp}=(4\pi n_{\rm imp}/3)^{1/3}$, where $n_{\rm imp}$ is the
number density of impurities. Electron-impurity scattering
is important at low temperatures, when electron-phonon scattering is
suppressed by quantum effects.

The expression for the frequency of electron-electron collisions
$\nu_{\rm ee}$ was obtained by Flowers and Itoh
\cite{Flowers&Itoh1976}. Their result can be written in the form
\begin{equation}
\label{Eq4}
    \nu_{\rm ee}=\frac{5 \pi^2 \alpha^2 k_{\rm B}^2 T^2}
    {2m_{\rm e}^\ast c^2\hbar}
    \left(\frac {k_{\rm F}}{k_{\rm TF}}\right)
    \left( 1+\frac 6{5x_{\rm r}^2}+\frac{2}{5x_{\rm r}^4}\right)
    \approx 4.473\cdot 10^{11}\left(\frac{k_{\rm F}}{k_{\rm TF}}\right)
    \,\left(\frac{n_0}{n_{\rm e}}\right)^{1/3}\,T_8^2 ~~\mbox{ s}^{-1},
\end{equation}
where the latter expression is presented for an ultra-relativistic
electron gas ($x_{\rm r}\gg 1$), $n_0\approx 0.16$~fermi$^{-3}$ is the
number density of nucleons in atomic nuclei, and $T_8$ is the
temperature in units of $10^8$ K.

%%%%%%%%%%%%%%%%%%%%%%%%%%%%%%%%%%%%%%%%%%%%%%%%%%
\subsection{The form factor of atomic nuclei}
%%%%%%%%%%%%%%%%%%%%%%%%%%%%%%%%%%%%%%%%%%%%%%%%%%

The function $u(q)$, which describes the Coulomb
electron-ion interaction
in Eq.~(\ref{Eq3}) has the form
\begin{equation}
    u(q)=\frac{F(q)}{q^2 |\varepsilon(q)|},
\end{equation}
where $\varepsilon(q)$ is the static longitudinal dielectric
function of the degenerate electron gas \cite{Jancovici1962}, which
takes into account the electron screening of the ion charge. Here,
\begin{equation}
\label{Eq5}
    F(q)\equiv\frac{1}{Z}\,\int e\,n_{\rm p}(\bm r)\,
        \exp(\imath\, {\bm r\cdot \bm q} )\,{\rm d} V
    =\frac{4\pi\, e}{Z}\, \int\limits_0^{r_{\rm p}}
        \frac{n_{\rm p}(r)\, \sin(qr)}{k}\,r\,{\rm d} r
\end{equation}
is the nuclear form factor that characterizes the distribution of proton
charge within an atomic nucleus. The integration in Eq.~(\ref{Eq5}) is
carried out over the atomic nucleus, $n_{\rm p}(r)$ is the local
number density of protons, and $r_{\rm p}$ is the radius of the
proton core. In white dwarfs and the outer envelopes of neutrons
stars ($\rho \lesssim 10^{11}$~g~cm$^{-3}$), the atomic nuclei can
be considered as pointlike, with $F(q)\equiv 1$. At densities
$\rho\lesssim 10^{13}$~g~cm$^{-3}$, the proton charge can (with good
accuracy) be taken uniformly distributed throughout the nucleus. In
this case, the form factor is
\begin{equation}
\label{Eq6}
    F(q)=\frac{3}{(qr_{\rm p})^3}
    \left[\sin(q r_{\rm p})-q r_{\rm p}\cos(q r_{\rm p})\right].
\end{equation}
%
%DY Already defined!!
%where $r_{\rm p}$ is the radius of the proton core in the atomic
%nucleus.
For $\rho\gtrsim 10^{13}$~{g$~$cm$^{-3}$}, the proton
density profile strongly differs from a step function, and the form
factor (\ref{Eq6}) becomes unacceptable. Then we have employed
the nuclear form factor provided by the model of the ground-state matter
(with the parameters smoothed over density jumps associated
with the changes of nuclides
 \cite{SmoothCompModel}).

%%%%%%%%%%%%%%%%%%%%%%%%%%%%%%%%%%%%%%%%%%%%%%%%%%%%%%%%%
\subsection{Analytic approximation for the viscosity}
%%%%%%%%%%%%%%%%%%%%%%%%%%%%%%%%%%%%%%%%%%%%%%%%%%%%%%%%%

We have obtained the analytic approximation for the Coulomb
logarithm of $ei$ scattering using the method of the {\it effective
electron-ion scattering potential} proposed in Ref.\ \cite{Potekhin1999}
for the electrical and thermal conductivities. The
conductivities of
the matter at $\rho\lesssim 10^{10}$~g~cm$^{-3}$ were
studied in \cite{Potekhin1999}, where the form factor of atomic
nuclei was taken to be unity. Later, Gnedin et al. \cite{Gnedin2000}
extended this method to higher densities using an approximate
nuclear form factor. As mentioned above, the effective structure
factors for the viscosity and electrical conductivity coincide. This
simplifies the generalization of the effective-potential method for
describing the shear viscosity. Following Ref.\ \cite{Gnedin2000},
we replace $u^2(q)S_\eta(q)$ in Eq.~(\ref{Eq3}) by
\begin{equation}
\label{Eq7}
    \left[ u^2(q)S_\eta(q)\right]_{\rm eff}
    =\frac{1}{\left(q^2+q_{\rm S}^2\right)^2}
    \left[1-{\rm e}^{-s_0q^2}\right]{\rm e}^{-s_1q^2}G_\eta D.
\end{equation}
The factor $\left(q^2+q_{\rm S}^2\right)^{-2}$ corresponds to the Debye
screening of the Coulomb interaction with the effective screening
length $q^{-1}_{\rm S}$; the term in the square brackets describes
ion correlations. The functions $G_\eta$ and $D$ characterize
ion quantum effects. The factor $\exp\left(-s_1q^2\right)$ added in
Ref.\ \cite{Gnedin2000} takes into account the effect of the
nuclear form factor. Numerical values of the shear
viscosity obtained from the exact theory are reproduced taking the same
parameters as for the electrical and thermal conductivities in
Ref.~\cite{Gnedin2000},
\begin{eqnarray}
    s&\equiv& \left(\frac{q_{\rm S}}{2k_{\rm F}}\right)^2
            =(s_{\rm i}+s_{\rm e})\,{\rm e}^{-\beta_Z};             \\
           \beta_Z &\!=\! & \pi\,\alpha\, Z\, \beta_{\rm r};\quad
            s_{\rm i}=s_{\rm D}\,(1+0.06\Gamma)\,
        {\rm e}^{-\sqrt \Gamma};\quad
            s_{\rm D}=(2k_{\rm F}\,r_{\rm D})^{-2};    \\
    w&\equiv& (2\,k_{\rm F})^2\,s_0=\frac{u_{-2}}{s_{\rm D}}\,
             \left(1+\frac{\beta_Z}{3}\right);            \\
    w_1&\equiv& (2\,k_{\rm F})^2\,s_1=14.73\,x_{\rm nuc}^2\,
        \left(1+\frac{Z}{13}\,\sqrt{x_{\rm nuc}}\right)
        \,\left(1+\frac{\beta_Z}{3}\right);                   \\
    G_\eta&=&\left(1+0.122\,\beta_Z^2\right)\,
             \left(1+0.0361\,\frac{Z^{-1/3}}{\Theta^2}\right)^{-1/2};        \\
    D&=&\exp\left[-0.42\,u_{-1}\,\sqrt{\frac{x_{\rm r}}{AZ}}
                \,\exp(-9.1\,\Theta)\right],
\end{eqnarray}
where
 $s_{\rm e}\equiv k_{\rm TF}^2/(2k_{\rm F})^2=\alpha /(\pi \beta_{\rm r})$
is the electron screening parameter,  $r_{\rm
D}=a/\sqrt{3\Gamma}$ is the ion Debye screening length,
 $x_{\rm nuc}$ is the ratio of the mean-square radius of
the proton distribution within an atomic nucleus to
the ion sphere radius; and $u_{-1}\approx 2.8$ and $u_{-2}\approx 13$
are the frequency moments of the phonon spectrum in the bcc Coulomb crystal.
Note that the function $G_\eta$ coincides with the function
$G_\sigma$ in Ref.~\cite{Gnedin2000}.

After integrating in Eq.~(\ref{Eq3}) with the effective potential
(\ref{Eq7}), we obtain
\begin{equation}
    \Lambda=\left[\Lambda_0(s,w+w_1)-\Lambda_0(s,w_1)\right]G_\eta D,
\end{equation}
where
\begin{equation}
    \Lambda_0(s,w)=\Lambda_1(s,w)-\left(1+\beta_{\rm r}^2\right)
    \Lambda_2(s,w)
    +\beta_{\rm r}^2\Lambda_3(s,w),
\end{equation}
\begin{eqnarray*}
    2\Lambda_1(s,w)&=&\ln\frac{s+1}s
        +\frac s{s+1}\,\left(1-{\rm e}^{-w}\right)
        -(1+sw)\,{\rm e}^{sw}\,\left[E_1(s\,w)-E_1(s\,w+w)\right],
    \\
    2\Lambda_2(s,w)&=&\frac{{\rm e}^{-w}-1+w}{w}
        -\frac{s^2}{s+1}\,\left(1-{\rm e}^{-w}\right)
        -2s\,\ln\frac{s+1}s \\
        &&+s\,(2+s\,w)\,{\rm e}^{sw}\,\left[E_1(sw)-E_1(sw+w)\right],
    \\
    2\Lambda_3(s,w)&=& 3\,s^2\,\ln  \frac {1+s}{s}
                    +\frac 12 \,\frac {2s^3-4s^2-3s+1}{1+s}
                    -\frac {s^3}{(1+s)}\,{\rm e}^{-w}
            +\frac{{\rm e}^{-w}}{w}
    \\
         &&+ \frac {(2sw-1)\left(1-{\rm e}^{-w}\right)}{w^2}
        -s^2 \,(3+sw)\, {\rm e}^{sw}\, \left( E_1(sw)-E_1 ( sw+w)\right).
\end{eqnarray*}
In this case, $E_1(x)\equiv\int_x^\infty y^{-1}{\rm e}^{-y}\,{\rm d} y$ is the
exponential integral (see, e.g., Ref.\ \cite{Abramovicz}). The
maximum error in our approximation of the viscosity does not exceed
20\%.

%%%%%%%%%%%%%%%%%%%%%%%%%%%%%%%%%%%%%%%%%%%%%%%%%%%%%%%
\subsection{The main properties of the shear viscosity}
%%%%%%%%%%%%%%%%%%%%%%%%%%%%%%%%%%%%%%%%%%%%%%%%%%%%%%%

\begin{figure}
    \begin{center}
        \leavevmode
        \epsfxsize=100mm \epsfbox[54 70 572 465]{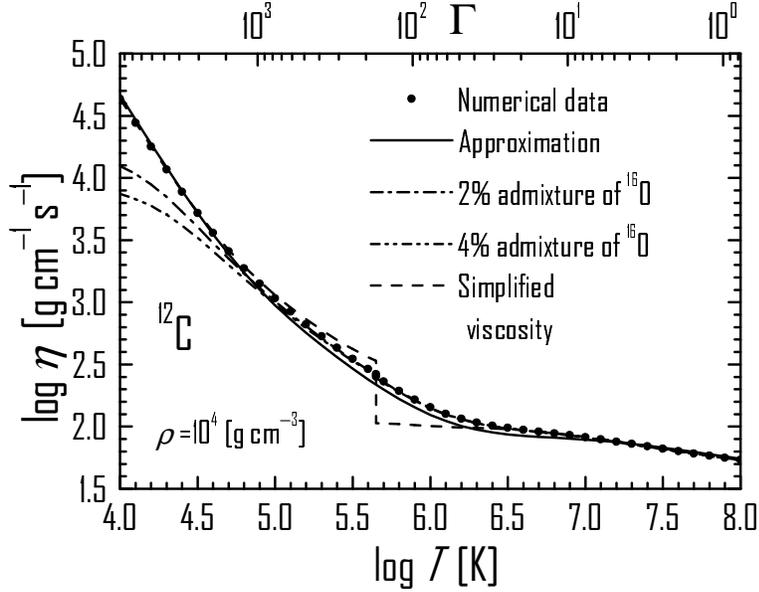}
    \end{center}
    %\includegraphics[viewport =28pt 48pt 544pt 460pt, width=100mm]{ChugFig1.eps}
    %\captionstyle{normal}
    \caption{Temperature dependence of the shear viscosity for
    a carbon plasma with the density $\rho = 10^4$\gsm. The solid curve is
    the analytic approximation. Bold points
    present numerical calculations. The dashed curve
    is the ''simplified'' viscosity, which demonstrates a jump at the
    melting point. The dot-dashed curves correspond to the matter with
    $^{16}$O impurities with concentration of 2 and 4\%.
     }
    \label{FigViscCarbon}
\end{figure}

Let us discuss the results of our calculations of the shear
viscosity neglecting the freezing out of Umklapp
processes (Section 2.2). Figure \ref{FigViscCarbon} presents the
temperature dependence of the shear viscosity for a carbon plasma
at $\rho = 10^4$~g~cm$^{-3}$. The upper horizontal axis
shows the Coulomb coupling parameter $\Gamma$ of ions. Because the
charge number is rather low, $Z = 6$, non-Born corrections are
small and invisible in Fig.\
\ref{FigViscCarbon}. All the data presented in the figure, except for
the dot-dashed curves, correspond to scattering of electrons by
ions of single type.

Bold points in the figure show numerical results. The solid
curve is the analytic approximation. The dashed
curve is the viscosity calculated using the simplified structure
factor (Section 2.2). Large jumps of this ''simplified''
viscosity at the melting point are clearly visible. These jumps (by
a factor of two to four) are present for all chemical elements and
all plasma parameters. Our modification of the structure factor (Section
2.2) increases the accuracy of calculations in the liquid and
solid phases, and nearly removes the viscosity jumps for all
elements. This makes it possible to introduce a unified approximation
for both phases (Section 2.4).

Nevertheless, appreciable viscosity jumps are still
present at the melting
point in our calculations for high densities, where zero point
oscillations of ions become important. We assume, as did Potekhin
{\it et al.} \cite{Potekhin1999} for the electrical and thermal
conductivities, that these jumps result from using the
classical structure factor in the ion liquid under the conditions when
quantum effects are important
(because quantum effects  are correctly included only in
the solid phase). Since numerical data used to construct the
analytic approximation include both the liquid and solid phases,
the unified analytic approximation shifts the viscosity in the
liquid phase to the viscosity in the solid phase. We expect that,
for an ion fluid at high densities, this approximation is more
exact than our original numerical data. It will be possible to
verify this in the future, when ion structure factors in a
fluid are calculated taking into account quantum effects.

The dot-dashed curves in Fig.\ \ref{FigViscCarbon} demonstrate the
effect of scattering by charged impurities. We have considered oxygen
impurities with concentrations of 2 and 4\%. The presence of these
impurities is weakly manifested at high temperatures, but dominates at
low temperatures, $T \ll T_{\rm p}$, when electron-phonon scattering
in the Coulomb crystal is strongly suppressed by quantum
effects.

\begin{figure}
    \begin{center}
        \leavevmode
        \epsfxsize=100mm \epsfbox[35 54 575 452]{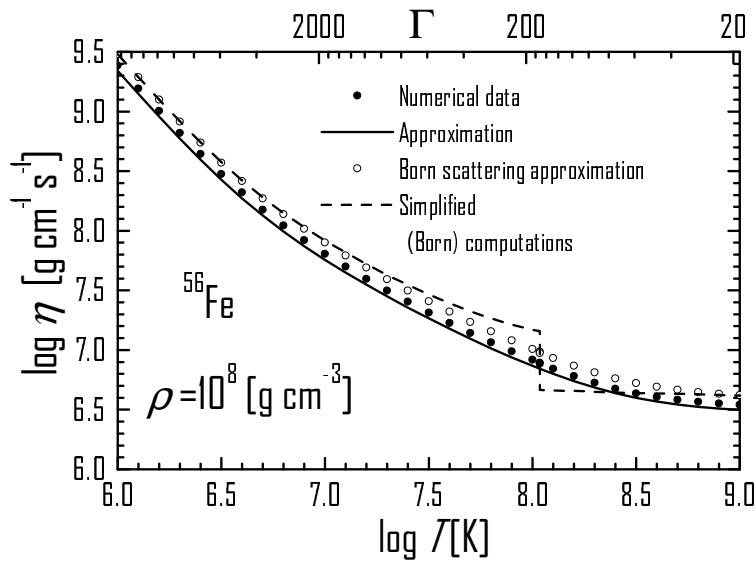}
    \end{center}
    \caption{Temperature dependence of the shear viscosity for
        an iron plasma with the density $\rho = 10^8$\gsm.
    The solid curve is the
        analytic approximation. Bold points present
        numerical calculations. Hollow circles
        correspond to the Born approximation. The dashed curve is the
        ''simplified'' viscosity calculated in the Born approximation.
    }
    \label{FigViscFerrum}
\end{figure}
Figure \ref{FigViscFerrum} shows the temperature dependence of
the shear viscosity for an iron plasma at $\rho =
10^8$\gsm. The upper horizontal scale plots the Coulomb coupling
parameter $\Gamma$. Bold points present our numerical results,
while the solid curve shows the analytic approximation. The dashed curve
is calculated with the simplified structure factor
neglecting non-Born corrections. As in Fig.\ \ref{FigViscCarbon},
the simplified viscosity displays jumps at the melting point, while
the new results pass smoothly through this point. The charge number
of iron ($Z = 26$) is high enough for the non-Born corrections to be
appreciable. To demonstrate this effect, hollow circles in Fig.\
\ref{FigViscFerrum} show the calculated
viscosity in the Born approximation. We can see that the
non-Born corrections reduce the viscosity by approximately 20\%.

\begin{figure}
    \begin{center}
        \leavevmode
        \epsfxsize=100mm \epsfbox[43 55 573 431]{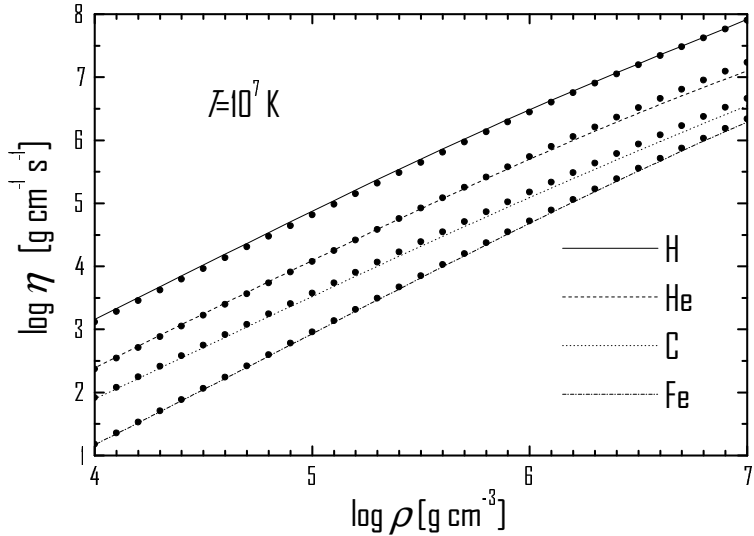}
    \end{center}
    \caption{Shear viscosity at the temperature $T =
        10^7$~K as a function of the density
    for various chemical compositions
        (H, He, C, and Fe). Bold points are the numerical results and
        the curves show the analytic fit.
    }
    \label{FigViskElems}
\end{figure}

Figure \ref{FigViskElems} shows the density dependence of the
shear viscosity for hydrogen, helium, carbon, and iron plasmas at
$T = 10^7$~K. We consider the densities typical
for the cores of white dwarfs and the outer envelopes of neutron
stars. Bold points show numerical results, and the curves
are analytic approximations. The strong dependence of the
viscosity on the chemical composition is due to the dependence of
an electron-ion collision frequency on the charge number $Z$.
In contrast to the thermal conductivity (see, e.g.,
Ref.\ \cite{Potekhin1999}), the effect of electron-electron collisions
on the viscosity is insignificant at the considered densities, even
for hydrogen.

\begin{figure}
    \begin{center}
        \leavevmode
        \epsfxsize=100mm \epsfbox[40 47 570 436]{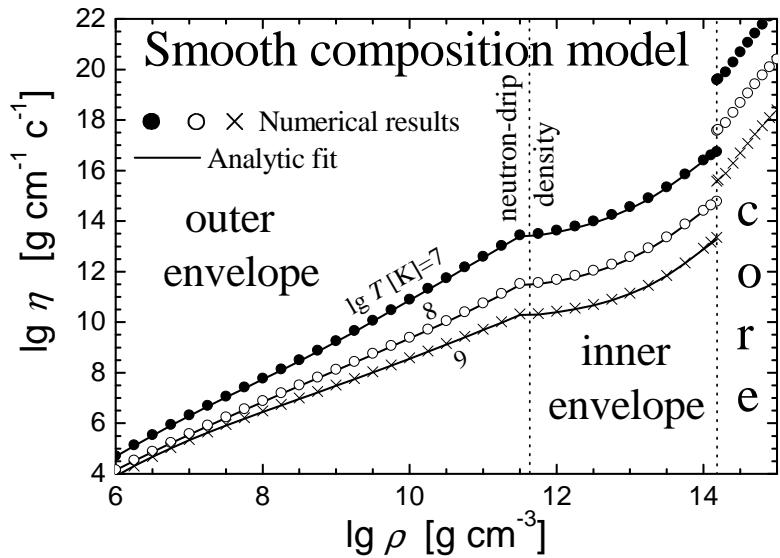}
    \end{center}
    \caption{Shear viscosity of the ground-state matter versus
    $\rho$ for the three temperatures $T = 10^7$, $10^8$,
    $10^9$~K. Solid curves show the analytic approximation;
   symbols are our numerical
    calculations. The vertical dotted lines indicate the neutron drip
    density and the crust-core interface in a neutron star
    (set to be $\rho = 1.5 \times 10^{14}$\gsm).
    The electron shear viscosity
    in the stellar core, determined by scattering of electrons by
    degenerate protons, is presented for comparison.
    }
    \label{FigViscSmooth}
\end{figure}

Figure \ref{FigViscSmooth} demonstrates the density dependence of
the shear viscosity in the density range from $10^6$ to
$10^{15}$\gsm for the three temperatures $T = 10^7$, $10^8$,
$10^9$~K. The ground-state nuclear composition is employed with
smoothed parameters. Symbols show the numerical results, and curves
are analytic approximations. In contrast to the thermal conductivity
\cite{Potekhin1999}, the shear viscosity decreases strongly with
growing temperature. Note that the ratio $\rho/\eta$ grows with
increasing density in the outer crust of a neutron star. These
results are important for the damping of oscillations in the neutron
star crust (see Section 3.3).

For illustrative purposes, the plot is continued beyond the crust
into the stellar core (to densities $\rho\ge 1.5 \cdot
10^{14}$\gsm). In the core, we have used the equation of state of the
matter presented in  Ref.~\cite{CoreEOS}. It is assumed that the core
is composed of neutrons, protons, and electrons and is not
superfluid. The electron viscosity in the core of such a star is
mainly determined by the scattering of electrons by
degenerate protons. The corresponding collision frequency is
obtained in the same way as the electron-electron
collision frequency (see Eq.~(\ref{Eq4})),
\begin{equation}
    \nu_{\rm ep}=\pi^2\alpha^2\,
        \left(\frac{k_{\rm F}}{q_0}\right)\,
        \frac{\left(k_{\rm B} T\right)^2\,{m_{\rm p}^\ast}^2}
            {\hbar\,p_{\rm F}^3}\, c
    \approx 1.434\cdot10^{12}\left(\frac{k_{\rm F}}{q_0}\right)\,
    T_8^2\left(\frac{m_{\rm p}^\ast}{m_{\rm p}}\right)^2\frac{n_0}{n_{\rm e}}\quad
    \mbox{c}^{-1},
\end{equation}
where $m_{\rm p}$ is the proton mass, and
$m^\ast_{\rm p}$ is its effective mass, which differs from $m_{\rm p}$
due to manybody effects
(we have set $m_{\rm p}^\ast=0.7\,
m_{\rm p}$). The Debye-screening parameter in the stellar core is
\begin{equation}
    q_0^2=4\pi\, \sum_j e_j^2\,\frac{\partial n_j}{\partial \mu_j},
\end{equation}
where the sum is taken over all types of charged particles
(electrons and protons); $e_j$, $n_j$ and $\mu_j$ are the charge,
number density, and chemical potential of particle species $j$. Owing
to a strong suppression of the scattering rate by proton degeneracy,
the electron viscosity in the core grows by a factor of $\sim 1000$
as compared to the viscosity in the crust.

%%%%%%%%%%%%%%%%%%%%%%%%%%%%%%%%%%%%%%%%%
\section{P mode oscillations of a neutron star crust}
%%%%%%%%%%%%%%%%%%%%%%%%%%%%%%%%%%%%%%%%%%%%%

This section is devoted to p mode
oscillations (i.e., oscillations in which perturbations of the
pressure dominate over the buoyant force) with high orbital numbers
(multipolarity), $l\gtrsim 500$, localized in the outer crust of a
nonrotating neutron star.

\subsection{General formalism}
\subsubsection{Flat metric for the envelope
     of a nonrotating neutron star}

The spacetime metric for a nonrotating neutron star
\cite{Shapiro&Tukolski1984} can be written as
\begin{equation}
\label{Eq8}
    {\rm d} s^2=c^2 {\rm e}^{2\Phi}\,{\rm d} \tilde{t}^2
               -{\rm e}^{2\lambda}\,{\rm d} r^2
               -r^2\,{\rm d}\Omega^2,
\end{equation}
where $\,{\rm d}\Omega^2=\,{\rm d}\theta^2 +\sin^2\,\theta\,{\rm
d}\varphi^2$, $\tilde{t}$ is the time coordinate, $r$ is the radial
coordinate, $\theta$ and $\varphi$ are the polar and azimuthal
angles, and the functions $\lambda(r)$ and $\Phi(r)$ determine
%%DY    %odetermine
spacetime curvature. It is sufficient to consider a thin envelope,
neglecting the variation of the functions $\lambda(r)$ and $\Phi(r)$
in the crust and using their values at the stellar surface,
\begin{equation}
    {\rm e}^{2\Phi(R)}={\rm e}^{-2\lambda(R)}
    =1-\frac{2GM}{c^2R},
\end{equation}
where $M$ is the gravitational stellar mass. Neglecting
variations of $r$ in the envelope compared to the stellar radius $R$
(in the approximation of a thin envelope layer), we can rewrite
Eq.~(\ref{Eq8}) in the form
\begin{equation}
    {\rm d} s^2=c^2\left(1-\frac{R_{\rm G}}{R}\right) \,{\rm d} \tilde{t}^2
               -\left(1-\frac{R_{\rm G}}{R}\right)^{-1}\,{\rm d} r^2
               -R^2\,{\rm d}\Omega^2,
\end{equation}
where $R_{\rm G}=2GM/c^2\approx 2.953\,(M/M_\odot)$~km is the
gravitational radius. Introducing the local time $t$ and the local
depth $z$, %specified by the relations
\begin{equation}
\label{Eq9}
    t= \tilde{t} \, \sqrt{1-R_{\rm G}/R},\quad
    z=(R-r)/\sqrt{1-R_{\rm G}/R},
\end{equation}
we come to a flat coordinate system in the
neutron star crust,
\begin{equation}
\label{Eq10}
 {\rm d} s^2=c^2\,{\rm d} t^2-\,{\rm d} z^2 -R^2\,{\rm d}\Omega^2.
\end{equation}
\subsubsection{Equilibrium structure of the crust}
The structure of a neutron star is determined by the equation of
hydrostatic equilibrium, including the effects of General Relativity
(the Tolman-Oppenheimer-Volkov equation; see, e.g.,
Ref.~\cite{Shapiro&Tukolski1984}). This equation is greatly simplified in
the envelope; in the coordinate system
(\ref{Eq9}) it can be written as
\begin{equation}
\label{Eq11}
    c_{\rm s}^2\frac{\,{\rm d}\rho_0}{{\rm d} z}=\frac{\,{\rm d}P_0}{{\rm d}z}
    =g\rho_0,
\end{equation}
where $P_0$ and $\rho_0$ are the equilibrium pressure and density,
$c_{\rm s}^2\equiv {\partial P_0}/{\partial \rho_0}$ is the square
of the local sound speed, and
\begin{equation}
    g=\frac{GM}{R^2\sqrt{1-R_{\rm G}/R}}
    \approx
      \,1.327\cdot 10^{14}\frac{M}{M_\odot}
      \left(\frac{10\,\mbox{km}}{R}\right)^2
      \left/\sqrt{1-R_{\rm G}/R}\right.
      \quad \frac{\mbox{cm}}{\mbox {s}^2}
\end{equation}
is the gravitational acceleration.

In computations we have used the equation of state of
a fully degenerate electron gas with electrostatic
corrections and have employed the model
of the ground-state matter with smoothed parameters.
Alternatively, we have used a polytropic envelope, composed of $^{56}$Fe
nuclei, where the pressure is determined by degenerate electrons,
assumed to be relativistic at all densities.

\subsubsection{The oscillation equation}

In the outer envelopes of neutrons stars, the main contribution to
the pressure is produced by degenerate electrons. Therefore,
while considering p-modes, we can use a
single equation of state to describe the equilibrium configuration
of the star and perturbations.

Let us use the Euler equation in a flat-space  metric (\ref{Eq10}),
\begin{equation}
    {\frac{\partial {\bm U}}{\partial {t}}}+({\bm U}\cdot\nabla){\bm U}
    =-\frac{\nabla  P}{\rho}+{\bm g},
\end{equation}
where $P$ is the pressure. The continuity equation
must also be satisfied,
\begin{equation}
    \frac{\partial \rho}{\partial t}
    +\nabla \left(\rho{\bm U}\right)=0.
\end{equation}
Assuming that the velocity $\bm U$ is small and introducing Euler
perturbations of the pressure $\delta P = P - P_0$  and the density
$\delta \rho =\rho - \rho_0$, we obtain the linearized Euler
equation
\begin{equation}
    {\frac{\partial {\bm U}}{\partial t}}=
    \frac{\delta\rho}{\rho_0^2}\,\nabla P_0
        -\frac 1{\rho_0}\,\nabla\delta P,
\end{equation}
and the continuity equation
\begin{equation}
\label{Eq12}
    \frac{\partial\delta \rho}{\partial t}
    +\nabla \left(\rho_0{\bm U}\right)=0,
\end{equation}
while the equation of state for the perturbations can be rewritten
in the form
\begin{equation}
\label{Eq13}
    \delta P=c_{\rm s}^2\delta\rho.
\end{equation}
We will consider irrotational motion and write the velocity in the
form ${\bm U}=\nabla \phi$, where $\phi$ is the velocity potential,
which is a scalar function of coordinates and time. Formally, the
function $\phi$ is determined up to an arbitrary function of time,
which we choose
in such a way that the Euler equation can be rewritten (using
Eqs.~(\ref{Eq11}) and (\ref{Eq13})) as
\begin{equation}
\label{Eq14}
    \frac{\partial\phi}{\partial t}=
    -\frac{\delta P}{\rho_0}=
    -c_{\rm s}^2\frac{\delta\rho}{\rho_0}.
\end{equation}
Differentiating (\ref{Eq14}) with respect to  $t$ and
taking into account Eqs.~(\ref{Eq12}) and (\ref{Eq11}), we get
\begin{equation}
\label{Eq15}
    \frac{\partial^2 \phi}{\partial t^2}=c_{\rm s}^2\Delta \phi
    +{\bm g}\cdot\nabla \phi,
\end{equation}
where we have introduced the Laplace operator
\begin{equation}
    \Delta \approx {\frac{\partial {^2}}{\partial {z^2}}} +\frac1{R^2}
    \left({\frac{\partial {^2}}{\partial {\theta^2}}}+
    \frac{1}{\sin^2\theta}{\frac{\partial {^2}}{\partial {\varphi^2}}}\right).
\end{equation}
An equation that coincides with (\ref{Eq15}) was obtained by Lamb
\cite{Lamb1911} for oscillations in the Earth atmosphere. The variables in
(\ref{Eq15}) can be separated if we write
\begin{equation}
    \phi={\rm e}^{\imath\omega t}\,Y_{lm}(\theta,\varphi)\,F(z),
\end{equation}
where $\omega$ is the oscillation frequency, $Y_{lm}(\theta,\varphi)$ is
a spherical harmonic function (see, e.g.,
Ref.~\cite{Varshalovich}), and $F(z)$ is an unknown function of the depth
determined by
\begin{equation}
\label{Eq16}
    {\frac{{\rm d} {^2F}}{{\rm d} {z^2}}}+\frac{g}{c_{\rm s}^2} {\frac{{\rm d} {F}}{{\rm d} {z}}}
    +\left(\frac{\omega^2}{c_{\rm s}^2}
    -\frac{l(l+1)}{R^2}\right)F=0.
\end{equation}
The first boundary condition for this equation,
\begin{equation}
\label{Eq17}
    \mbox{function } F(z)\mbox{ is bounded as }z\rightarrow0,
\end{equation}
follows from the requirement of a finite
oscillation amplitude
at the stellar surface. The second boundary condition is
imposed artificially. In the current study, we have to solve equations that
are applicable only in the thin stellar crust. Therefore,
oscillations should be damped at sufficiently
high depths. For simplicity, we
formally move this boundary condition to infinity along $z$ and will
check a true localization of oscillations (see
Section 3.3). In this case, the second boundary condition can be written
\begin{equation}
\label{Eq18}
    F(z)\rightarrow 0~\mbox{ as }z\rightarrow \infty.
\end{equation}
Together with the boundary conditions (\ref{Eq17}) and (\ref{Eq18}),
Eq.\ (\ref{Eq16}) specifies eigenfrequencies and eigenmodes of
oscillations. Moreover, the following asymptotes are valid at large
and small depths:
\begin{equation}
\label{Eq19}
    F(z)\propto \left\{
    \begin{array}{ll}
        1-\omega^2z/g \quad &z\rightarrow 0,
        \\
        \exp\left(-\sqrt{l(l+1)}\,z/R\right)\quad&
                z\rightarrow \infty.
    \end{array}
    \right.
\end{equation}
Perturbations of the pressure and the density are expressed in terms of
the function $\phi(r)$ using Eq.~(\ref{Eq14}),
\begin{equation}
    \delta P=-\imath\omega\,\rho_0\,\phi,\quad
    \delta \rho=\frac{\delta P}{c_{\rm s}^2}
               =-\imath\frac{\omega\,\rho_0}{c_{\rm s}^2}\,\phi.
\end{equation}
Due to the boundary condition (\ref{Eq17}), variations of the
pressure and the density, $\delta P$ and $\delta\rho$, vanish at the
stellar surface (because $\rho_0(R) = 0$). We can see from these last
expressions that the number of radial nodes ($k$)
of the velocity potential
coincides with the number of nodes of the pressure and
density variations. Below, we will call $k$ the {\it number of
radial nodes of the mode}.

The displacement vector for an oscillating matter element
can be written in the form
\begin{equation}
    {\bm \xi}\equiv\int {\bm U}\, {\rm d}t
    =-\frac{\imath}{\omega}\,\nabla\cdot\phi.
\end{equation}
The z component of this vector is
\begin{equation}
    \xi_z=-\frac{\imath}{\omega}\,Y_{lm}(\theta,\varphi)\,{\frac{{\rm d} {F}}{{\rm d} {z}}},
\end{equation}
and the magnitude of horizontal displacement can be estimated as
\begin{equation}
    \left|\xi_{\rm h}\right|\approx \frac{l}{\omega\,R}\,\left|F(z)\right|.
\end{equation}
The quantities  $l F(z)/R$ and ${\rm d} F/{\rm d} z$
appear in Eq.~(\ref{Eq16}) on equal
footing. Therefore, horizontal and radial
displacements should have the same order of magnitude for the
oscillations considered.

Oscillations of a polytropic envelope in a plane-parallel
approximation were studied earlier by Gough \cite{Gough} assuming
that the equations of state of the unperturbed matter and the
perturbations are polytropic with different indices. In
the limiting case of the same polytropic index $n$, his result can be
presented as follows. The mode containing $k$ radial nodes has the
eigenfrequency
\begin{equation}
\label{Eq20}
    \omega^2_k=\frac gR\,
    \sqrt{l(l+1)}\,\left(\frac{2k}n+1\right)\,
    \approx 10^8 \,g_{14}\,\left(\frac{10\,\mbox{km}}{R}\right)
   \,\sqrt{l(l+1)}\,\left(\frac{2k}n+1\right)~~\mbox{s}^{-2},
\end{equation}
while the velocity potential is given by
\begin{equation}
     F_k(z)=\exp\left(-\sqrt{l(l+1)}\,\frac{z}{R}\right)
            L_k^{(n-1)}\left(2\,\sqrt{l(l+1)}\,\frac{z}{R}\right),
\end{equation}
where  $L_k^{(n-1)}(x)$ is a generalized Laguerre polynomial (see,
e.g., Ref.~\cite{Abramovicz}) and $g_{14}$ is the gravitational
acceleration at the stellar surface in units of
$10^{14}$~cm~s$^{-2}$. Note that the eigenfrequencies agree with the
simple estimate $\omega^2\sim g/a$, where $a\sim R/l$ is the
characteristic depth for the oscillation localization.

Let us remark that the mode with $k = 0$,
which does not have any radial nodes,
corresponds to the vanishing Lagrangian variations of the pressure
and the density ($\nabla \cdot {\bm U}$=0, see (\ref{EqA2})); its
parameters do not depend on the polytropic index. Adding the
condition $\nabla \cdot {\bm U}\equiv \Delta \phi=0$ to
Eq.~(\ref{Eq15}), it is easy to show that this mode is described
by the function $F(z)=\exp\left(-\sqrt{l(l+1)}z\,/R\right)$ and has
the frequency
\begin{equation}
\label{Eq21}
    \omega_0^2=\frac gR \, \sqrt{l(l+1)}
    \approx10^8 \,g_{14}\,\left(\frac{10\,\mbox{km}}{R}\right)
    \,\sqrt{l(l+1)}
    ~~\mbox {s}^{-2};
\end{equation}
it exists for any equation of state. Therefore, the frequency
$\omega_0$ will further be used to normalize  oscillation
frequencies.

The frequencies $\omega$ we discuss refer to the coordinate
system of the stellar envelope (see Eq.~(\ref{Eq9})). They are easily
transformed to the frequencies $\widetilde\omega$ detected by a
distant observer,
\begin{equation}
       \widetilde\omega=\omega \, \sqrt{1-R_{\rm G}/R} .
\end{equation}
%

%%%%%%%%%%%%%%%%%%%%%%%%%%%%%%%%%%%%%%%%%%%%%%%%
\subsection{Viscous damping of oscillations}
%%%%%%%%%%%%%%%%%%%%%%%%%%%%%%%%%%%%%%%%%%%%%%%%

 In this section, we consider the damping of oscillations
with velocity potentials of the form ${\rm e}^{\imath \omega
t}\,Y_{lm}(\theta,\varphi)F(z)$ in a spherically symmetric star
under the action of shear viscosity. We take flat space-time metric
which is applicable to oscillations studied in Section 3.1. This is
because the flat metric (\ref{Eq10}), which coincides with the
metric for a thin spherical layer in a flat space-time, can be
introduced in the region, where oscillations are localized. As a
result, it is sufficient to consider the oscillation damping time in
a flat metric and transform this time (in accordance with
Eq.~(\ref{Eq9})) into the frame of a distant observer. We define the
oscillation damping time $\tau$ as

\begin{equation}
\label{Eq22}
    \tau=E / |{\rm d}E/{\rm d}t|,
\end{equation}
where
\begin{equation}
\label{Eq23}
    E=\int
%    \left(\rho\frac{|U|^2}{4} +\frac{c_{\rm s}^2|\delta\rho|^2}{4\rho_0}\right)
      \varepsilon
      \,{\rm d} V
     =\int \rho \,\frac{|U|^2}{2}\,{\rm d} V
\end{equation}
is the total energy of oscillations and
\begin{equation}
    \varepsilon=\frac 14 \left( \rho_0\,|U|^2
      +\frac{c_{\rm s}^2}{\rho_0}\,|\delta\rho|^2\right)
\end{equation}
is the energy density of oscillations at a given point averaged
over an oscillation period (see, e.g., Ref.~\cite{Gidrodinamika}). The
additional factor of $1/2$ in the expression for $\varepsilon$ is
required due to the averaging over the oscillation period. The
integration is carried out over the entire volume of the star (in
practice, over the region where the oscillations are localized). We
neglect perturbations of the gravitational potential. The last
equality in Eq.~(\ref{Eq23}) is determined by the equality of the mean
kinetic and potential energies for low-amplitude harmonic
oscillations. Note that some authors introduce the
damping time for the oscillation amplitude rather than the damping
time (\ref{Eq22}) for the oscillation energy.

When calculating the energy from Eq.~(\ref{Eq23}), the angular
integration can be carried out analytically,
\begin{equation}
    E=\frac 12 \int_0^{R} \rho \left[\left(F\,^\prime\right)^2
         +\frac{l(l+1)}{r^2}F^2\right] r^2\,{\rm d}r.
\end{equation}
The period-averaged rate of the viscous energy dissipation is
(see, e.g., Ref.~\cite{Gidrodinamika})
\begin{equation}
\label{Eq24}
    \frac {\,{\rm d}E}{\,{\rm d}t}=-\frac 14 \int
    \sigma_{ik}^\prime
          \left(\frac{\partial U_i^\ast}{\partial x_k}
                +\frac{\partial U_k^\ast}{\partial x_i}
          \right)
    \,{\rm d} V,
\end{equation}
where the viscous stress tensor $\sigma_{\alpha\beta}^\prime$ is
given by Eq.~(\ref{Eq1}). As in the expression for the oscillation
energy $E$, the additional factor of $1/2$ is required owing to
averaging over the oscillation period. It is easy to see that
the energy dissipation rate separates into a sum of terms
associated with shear and bulk viscosities. We will restrict ourselves
to the dissipation determined by the shear viscosity. The integration
over the angular variables in Eq.~(\ref{Eq24}) can be carried out
analytically (see the Appendix).

\subsection{Discussion of numerical results}

As an example, let us choose a ''canonical'' model for a neutron star
with the mass $M = 1.4M_\odot$ and the radius $R = 10$~km. For
this model,
\begin{equation}
    \omega_0\approx 1.56 \cdot 10^5 \left(\frac{l(l+1)}{10^4}\right)^{1/4}
    \,\mbox{s}^{-1}
\end{equation}
and for a distant observer
\begin{equation}
    \widetilde{\omega}_0\approx 0.766\,\omega_0\approx 1.19 \cdot 10^5
    \left(\frac{l(l+1)}{10^4}\right)^{1/4}\mbox{~s}^{-1}.
\end{equation}
The thickness of the outer crust in such a star ($\rho < 4 \cdot
10^{11}$ g$\cdot$cm$^{-3}$, before the neutron drip point) is
$\approx 364$~m.

Oscillation eigenfrequencies have been found via a series of
iterative trials, checking for the coincidence of a mode number
with a number of radial nodes.

%----------------------------------------------
\subsubsection{Oscillation eigenfrequencies}
%----------------------------------------------

%
\begin{figure}
    \begin{center}
        \leavevmode
        \epsfxsize=100mm \epsfbox[28 130 567 440]{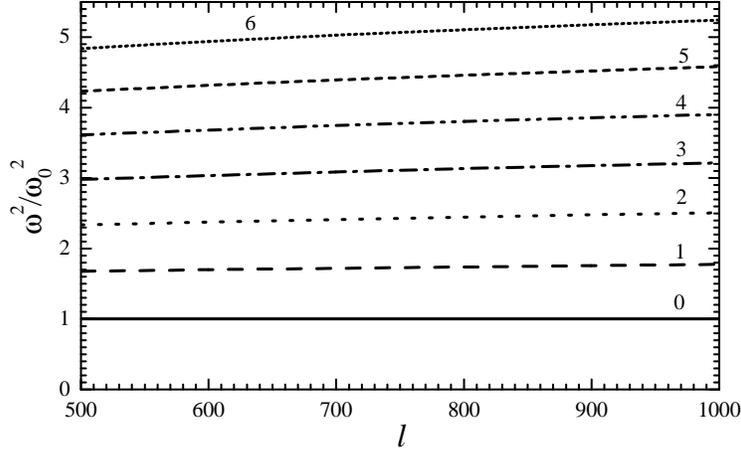}
    \end{center}
    \caption{Eigenfrequencies of oscillations localized in the
            crust of a ''canonical'' neutron star. The frequencies are
            normalized to the frequency $\omega_0$ given by Eq.~(\ref{Eq21}).
        The numbers next to
            the curves indicate the number of radial nodes.
    }
    \label{FigOmegas}
\end{figure}
\begin{figure}
    \begin{center}
        \leavevmode
        \epsfxsize=100mm \epsfbox[21 130 530 420]{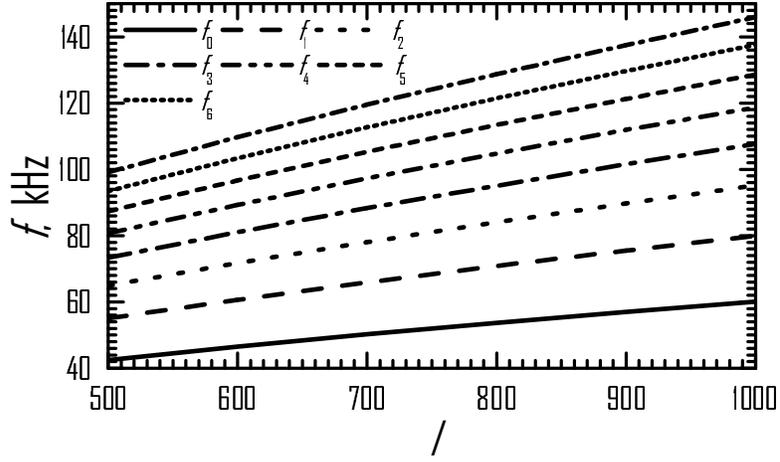}
    \end{center}
    \caption{Frequencies of oscillations localized in the crust
         of a ''canonical'' neutron star as detected by a distant observer.
     The
         subscript of $f$ denotes the number of radial nodes.
    }
    \label{FigOmegasValues}
\end{figure}

The dependence of oscillation eigenfrequencies on $l$,
calculated from Eq.~(\ref{Eq16}) with the boundary conditions
(\ref{Eq17}) and (\ref{Eq18}), is presented in Figs.\
\ref{FigOmegas} and \ref{FigOmegasValues}. As mentioned earlier,
the frequency of the fundamental mode, which does not have any
radial nodes, is determined by Eq.~(\ref{Eq21}) for all $l$. With
decreasing $l$, oscillations penetrate deeper into the
outer crust, where the equation of state is softened because
the electron gas becomes relativistic and electrons undergo
beta captures. This slightly decreases
the dimensionless eigenfrequencies.
As in the model with the polytropic equation of state
(see Eq.~(\ref{Eq20})), the separation between squares of
neighboring eigenfrequencies for any $l$ is nearly constant.
A weak decrease of this separation
with the growing number of radial nodes is due to the
penetration of oscillations into deeper layers of the star, where the
equation of state is softer. When $l\sim 500$, the main oscillation
energy is localized in the region 50~m$\lesssim\, z\, \lesssim
$\,400~m, where the equation of state of the degenerate,
relativistic electron gas well describes  the profile of the sound
speed. Therefore, the estimate (\ref{Eq20}) for eigenfrequencies
of the polytropic crust model  with the
polytropic index $n = 3$ is justified (deviations constitute
several per cent).

\subsubsection{Oscillation eigenmodes}
\begin{figure}
    \begin{center}
        \leavevmode
        \epsfxsize=100mm \epsfbox[23 120 563 456]{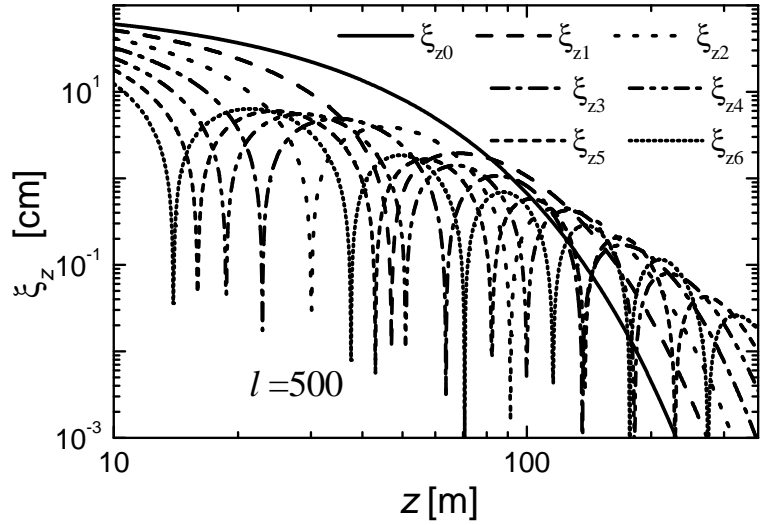}
    \end{center}
    \caption{Root-mean-square (over angles) amplitude
        of radial displacements of matter for modes with $l = 500$. The
        subscript of $\xi$ indicates the number of radial nodes. The
        root-mean-square amplitude of radial displacements at the
        stellar surface has been set equal to 1~m.}
    \label{FigXiZ_l500}
\end{figure}

Figure \ref{FigXiZ_l500} presents profiles of radial
displacements of matter elements for the modes with $l = 500$.
The root-mean-square amplitude of radial displacements of the stellar surface
has been set 1~m. Since we are studying linear oscillations,
this quantity is an arbitrary (sufficiently small) constant that
normalizes the solution. From Fig.\ \ref{FigXiZ_l500} one can easily
determine  the magnitude of radial displacements in the star for
any other amplitude of the displacements at the surface. With
growing $z$, the
radial displacement amplitude $\xi_z$ overall decreases. At
$z\sim 300$~m, the decrease becomes monotonic and gradually tends to
the exponential asymptotic (\ref{Eq19}). This means that
oscillations are localized in the outer crust.

\begin{figure}[h!t]
    \begin{center}
        \leavevmode
        \epsfxsize=100mm \epsfbox[18 120 569 410]{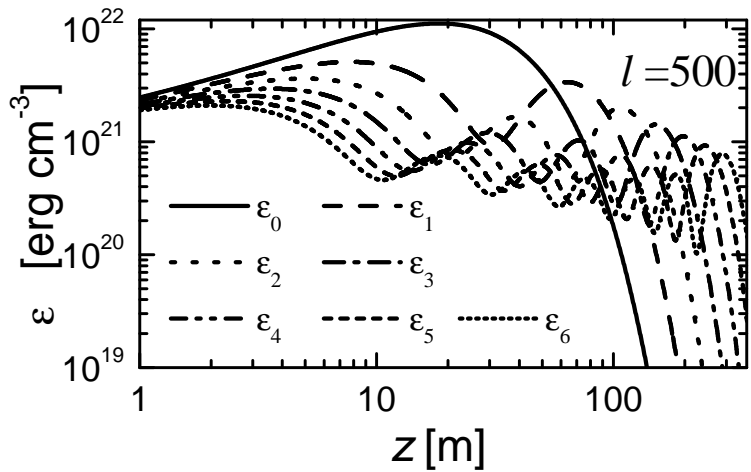}
    \end{center}
    \caption{Angle-averaged energy density $\varepsilon$ of oscillations
    with $l = 500$. The subscript of $\varepsilon$
    indicates the number of
    radial nodes. The root-mean-square amplitude of radial displacements
    at the stellar surface is 1~m.
    }
    \label{FigEtot_l500}
\end{figure}

This effect is manifested even more clearly in the energy density of
oscillations. Figure \ref{FigEtot_l500} presents the dependence
of the angle-averaged energy density $\varepsilon$
versus depth $z$ for oscillations with $l =
500$. The mode amplitude is normalized in the same way as in
Fig.\ \ref{FigXiZ_l500}. In our approximation, the energy density
is proportional to the square of the normalized
amplitude of radial surface displacements. The depicted
modes are localized in the outer crust of the star. The
energy density of oscillations varies relatively weakly within
the ''critical'' depth $z\lesssim 100-200$~m, after which it falls
off exponentially. The energy density decreases by more than two
orders of magnitude near the boundary between the outer and inner
crusts.

\begin{figure}[h!t]
    \begin{center}
        \leavevmode
        \epsfxsize=100mm \epsfbox[20 110 547 390]{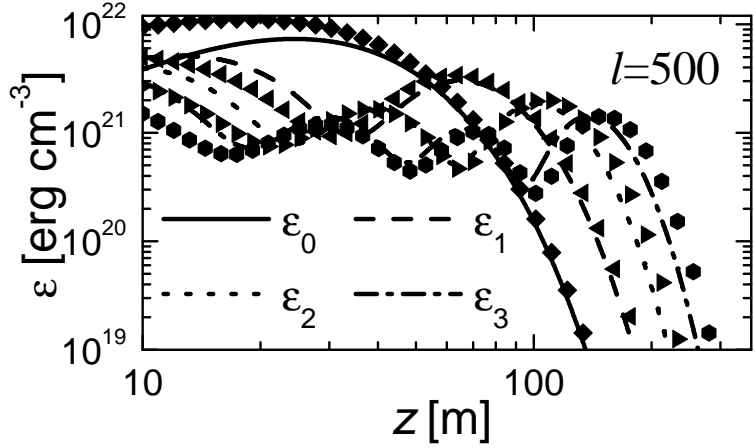}
    \end{center}
    \caption{Angle-averaged energy density $\varepsilon$
    of oscillations with $l = 500$.
    The subscript of  $\varepsilon$ indicates the
    number of radial nodes. The curves refer to the
    polytropic model, and the symbols are for the
    exact equation of state.}
    \label{FigEtot_l500_Model}
\end{figure}
As noted in Section 3.3.1, at $l\sim 500$  oscillation
frequencies are well reproduced by a polytropic crust model.
The situation is somewhat different for oscillation eigenmodes.
Normalization at the stellar surface is not expedient for these
modes, since the polytropic model
poorly reproduces the structure of the star
at low depths, $z \lesssim 40$~m. Consequently, such a normalization
leads to large errors at the depths of interest to us, $z\lesssim
100-200$~m, where the main oscillation energy is concentrated.
Therefore, we need a special normalization to compare
exact and polytropic modes. Figure
\ref{FigEtot_l500_Model}
(similar to Fig.\ \ref{FigEtot_l500}) shows the
angle-averaged energy density $\varepsilon$ of oscillations as a
function of $z$. The symbols refer to exact numerical results,
while the curves are for the polytropic
model, normalized in such a way to be in agreement with the exact
results
in the region where oscillations are localized. We can conclude that the
polytropic model for the outer crust satisfactorily reproduces the
energy density of oscillations at depths of 60~m\,$\lesssim\,
z\, \lesssim\, 300$~m for modes with $l\lesssim 500$.

\subsubsection{The oscillation damping}
\begin{figure}[th]
    \begin{center}
        \leavevmode
        \epsfxsize=100mm \epsfbox[44 85 575 385]{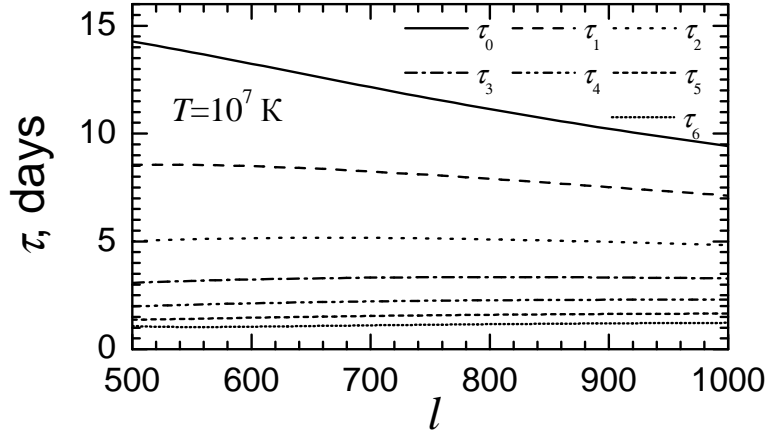}
    \end{center}
    \caption{Damping time $\tau$ of oscillations for a distant
    observer as a function of multipolarity  $l$
    for a neutron star with the internal
    temperature $T = 10^7$~K. The subscript of $\tau$ gives the
    number of radial nodes.
    }
    \label{FigZatuh_T7}
\end{figure}
\begin{figure}[ht]
    \begin{center}
        \leavevmode
        \epsfxsize=100mm \epsfbox[44 85 575 385]{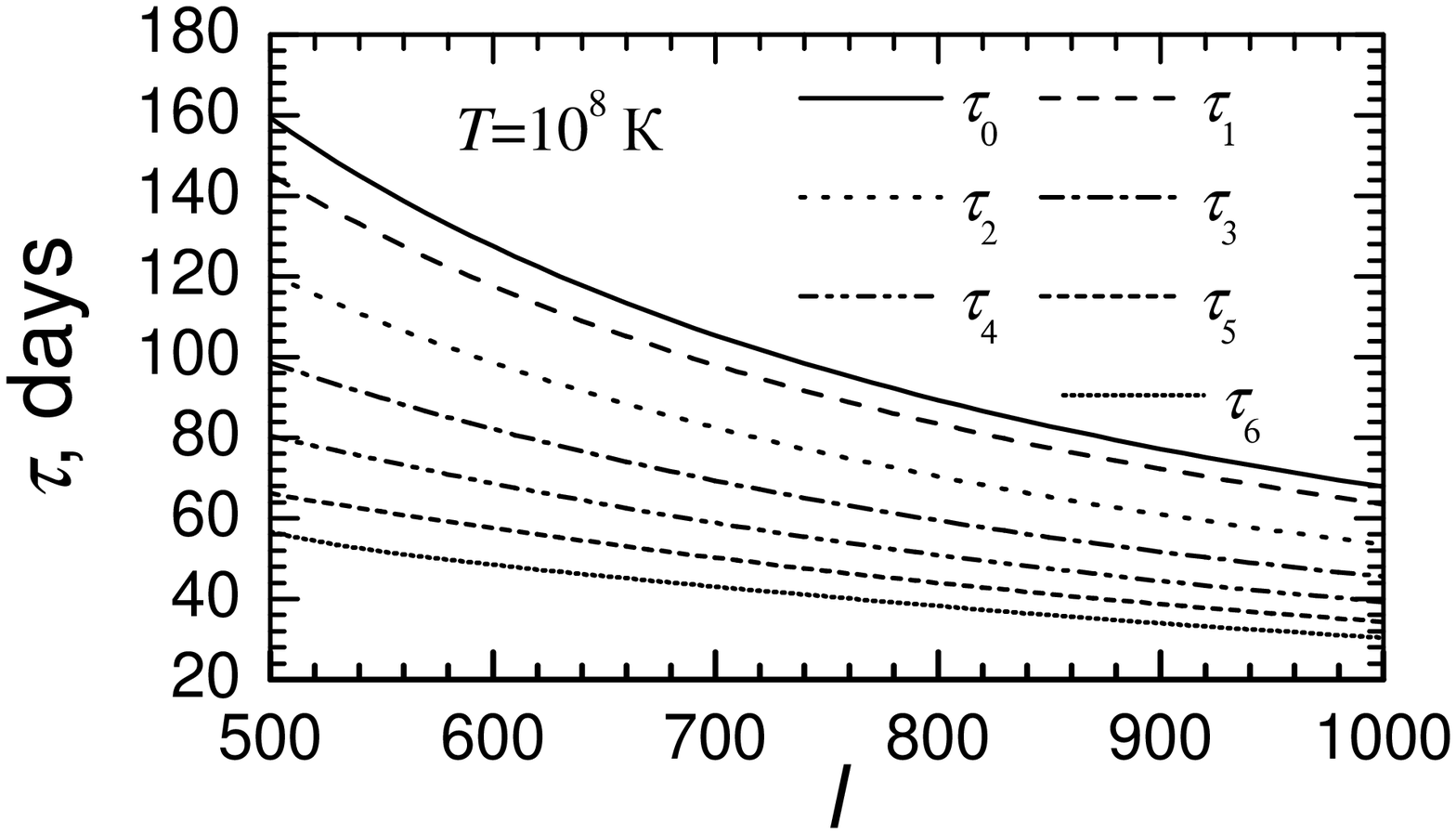}
    \end{center}
    \caption{Same as Fig.\ \ref{FigZatuh_T7} but
    for the internal temperature $T=10^8$~K.
    }
    \label{FigZatuh_T8}
\end{figure}
\begin{figure}[ht]
    \begin{center}
        \leavevmode
        \epsfxsize=100mm \epsfbox[44 85 575 385]{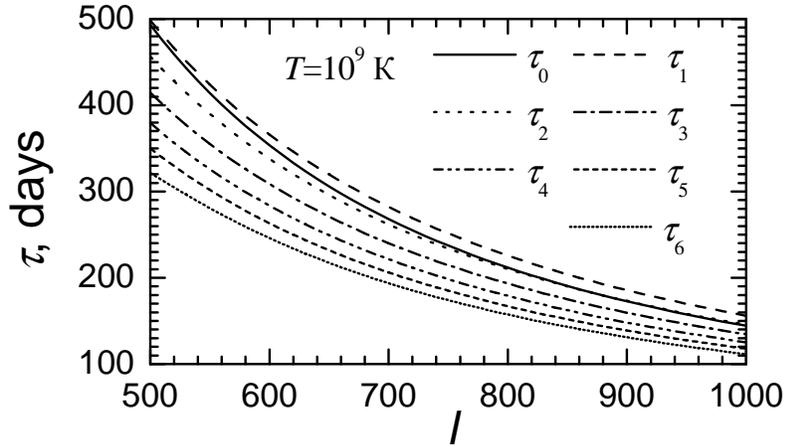}
    \end{center}
    \caption{Same as Fig. \ref{FigZatuh_T7} but for
    the internal temperature $T=10^9$~K.
    }
    \label{FigZatuh_T9}
\end{figure}

In our calculations we have considered the neutron star crust as
isothermal. This approximation describes well the internal
temperature profile. The temperature is nearly independent of depth
due to a high thermal conductivity of degenerate electrons.
Naturally, oscillation frequencies and damping times
do not depend on the normalization amplitude (the
amplitude of radial surface displacements). Figures
\ref{FigZatuh_T7}, \ref{FigZatuh_T8} and \ref{FigZatuh_T9} present
the dependence of the oscillation damping time $\tau$
(for a
distant observer) on $l$ for a canonical neutron star with the internal
temperature $T = 10^7,$ $10^8$, and $10^9$~K. The strong
temperature dependence of the damping time is due to the
appreciable decrease of the viscosity with increasing temperature
(Fig.\ \ref{FigViscSmooth}). The damping time can be
estimated from characteristic oscillation parameters  as
\begin{equation}
    \tau\sim \varepsilon/{\dot{\varepsilon}}\sim\rho U^2\left/
    \left(\eta\left(\frac{U}{\lambda}\right)^2\right)\right.
    \sim
    \lambda^2\frac{\rho}{\eta},
\end{equation}
where $\dot{\varepsilon}$ is the local viscous dissipation rate, while
$U$ and $\lambda$ are, respectively,
the characteristic velocity and its variation length scale
in the region of oscillation localization.

Let us consider Fig.\ \ref{FigZatuh_T8} in more detail. Oscillations
with $l\sim 500$ are localized at $z\lesssim 100$~m (Fig.\
\ref{FigEtot_l500}), which corresponds to densities $\rho\lesssim
10^{10}$\gsm. Under these conditions, the ratio $\rho/\eta$ is $\sim
3$~s~cm$^{-2}$ (Fig.\ \ref{FigViscSmooth}) and grows with increasing
$l$ (due to the decrease of the density in the region of oscillation
localization). Let us make estimates for the modes with $l\sim 500$.
The velocity variation length scale can be estimated as $\lambda
\sim R/l$. Note that this scale decreases for modes with a large
number of radial nodes, accelerating the damping of these
oscillations. For the fundamental mode (without any radial nodes),
the damping time (in the frame of a distant observer) can be
estimated as
\begin{equation}
     \tau\sim 1.2\,10^{4}\,\times(500/l)^2\mbox{ s}\approx
     120\times
     (500/l)^2\mbox{~day},
\end{equation}
in good agreement with the numerical results for $l\sim 500$. The
damping time drops weaker than $\propto l^{-2}$ because
the ratio $\rho/\eta$ grows for higher $l$, owing to the decrease of
the density in the region of oscillation localization.

For the outer crust of a star with the temperature $T = 10^9$~K,
the ratio $\rho/\eta$ depends weakly on $\rho$.
Therefore, the oscillation damping time obeys the law $\tau
\propto l^{-2}$.

\begin{figure}[t!h]
    \begin{center}
        \leavevmode
        \epsfxsize=100mm \epsfbox[44 120 575 400]{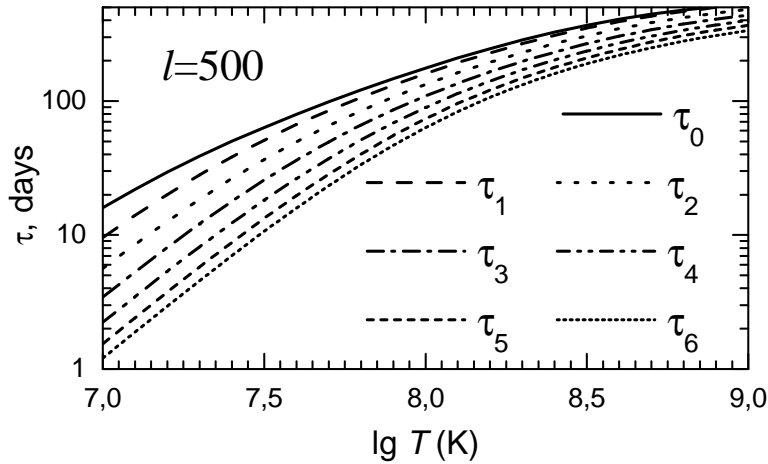}
    \end{center}
    \caption{Damping time of oscillations as a function
    of the internal neutron star temperature for modes with $l = 500$.
    The subscript indicates the number of radial nodes. Numerical
    results are redshifted for a distant observer.}
    \label{FigZatuhT}
\end{figure}

The damping time as function of the crust temperature is presented
in Fig.\ \ref{FigZatuhT}, where we have chosen the modes with $l = 500$
as an example. The damping time grows by approximately two orders of
magnitude as the temperature varies from $10^7$ to $10^8$~K. When
the temperature increases by another order of magnitude, the damping
time grows further by a factor of three. This is due to the
nonlinear decrease of the viscosity with the growth of $T$
(Fig.\ \ref{FigViscSmooth}).

There are many other oscillation damping mechanisms \cite{McDermott1988},
in addition to the viscous
damping considered here. For example,
one often considers the damping due to the emission of
gravitational and electromagnetic waves (for oscillations of the
stellar matter with a frozen-in magnetic field).
In our case, these mechanisms are inefficient due to high
multipolarity, $l\sim 500$. If $l$ is high, we expect
gravitational or electromagnetic radiation to be generated by an
ensemble of closely spaced coherent elementary radiating regions,
which radiate in antiphase and cancel one another. Formally, the
weakness of this radiation is manifested by the presence of large
factors $(2l +1)!!$ in the denominators of the expressions for the
radiation intensities (see, e.g., Refs.\ \cite{bal82,McDerm84}).
An analysis shows that the damping of oscillations we consider here
is determined to a great extent by the shear viscosity.

A detailed analysis of the evolution of pulse shapes of some
radio pulsars provides an evidence that high-multipole oscillations
are, indeed, excited in them (see, e.g., the recent paper
\cite{clemens04}). However, reliable observational data on the
existence of such oscillations have not yet been obtained.

%%%%%%%%%%%%%%%%%%%%%%%%%%%%%%%%%%%%%%%%%%%%%%%%%%%%%%%%%%
\section{Conclusions}
%%%%%%%%%%%%%%%%%%%%%%%%%%%%%%%%%%%%%%%%%%%%%%%%%%%%%%%%%%

We have calculated the shear viscosity of dense stellar matter for a
broad range of temperatures and densities typical for the cores of
white dwarfs and the envelopes of neutron stars. We have considered
the matter composed of astrophysically important elements, from H to
Fe, at the densities from $10^2-10^4$\gsm to $10^7-10^{10}$\gsm. At
higher densities, $10^{10}-10^{14}$\gsm, we used the model for the
cold ground-state matter taking into account finite sizes of atomic
nuclei and the distribution of proton charge within the nuclei.
Under the conditions described above, the shear viscosity is
determined by Coulomb scattering of degenerate electrons by atomic
nuclei. We have used the modified structure factor of ions proposed
by Baiko {\it et al.} \cite{Baiko1998} and applied by Potekhin {\it
et al.} \cite{Potekhin1999} for calculating the thermal and
electrical conductivities of dense matter. In the ion liquid, this
modification approximately takes into account a quasi-ordering in
ions positions, which reduces the electron-ion scattering rate. In
the crystalline phase, the new structure factor takes into account
multi-phonon processes, which are important near the melting
temperature $T_m$. The new results near the melting point differ
appreciably from the results of Flowers and Itoh
\cite{Flowers&Itoh1976,Flowers&Itoh1979} obtained for a Coulomb
liquid. We have approximated our numerical results by an analytic
expression convenient for applications.

We have analyzed eigenfrequencies and
eigenmodes of oscillations localized
in the outer crust of a neutron star (in the plane-parallel
approximation). A polytropic model for the crust
can reasonably well reproduce the
eigenfrequencies of oscillation modes with multipolarity $l\sim
500$. We have also calculated viscous damping times of the oscillations.
We have obtained a sharp decrease in the damping time
with decreasing the internal temperature of a neutron star. For example,
for a neutron star with mass $M = 1.4M_\odot$, radius $R = 10$~km,
and the internal temperature $T = 10^8$~K, the damping time for the
fundamental mode with $l = 500$ is $\sim 160$~days. When the
temperature decreases to $T\sim 10^7$~K, the damping time falls to
$\sim 15$~days.

In our calculations, we have used the model of the ground-state matter in
the neutron star crust with a smoothed dependence of the parameters
of atomic nuclei on the density of the matter. More accurate
calculations would  require a more detailed model of the
ground-state matter, where the composition varies with
depth in a jump-like manner (showing a series of weak first-order phase
transitions at some depths). The
presence of these jumps could amplify the damping of the
oscillations. We plan to consider this problem in the future.

%%%%%%%%%%%%%%%%%%%%%
\begin{center}
    {Acknowledgments}
\end{center}
%%%%%%%%%%%%%%%%%%%%%%

We thank D.\ Gough for reprints of his papers on
oscillations of polytropic stellar envelopes, and W.\ Dziembowski,
who pointed out these works to us. We are also grateful to K.P.\
Levenfish and A.Y.\ Potekhin for useful discussions. This work was
supported by a student grant of the Non-Profit Foundation
''Dynasty'' and the International Center for Fundamental Physics in
Moscow, the Russian Foundation for Basic Research (RFFI-IAU grant
no.\ 03-02-06803 and project no.\ 05-02-16245), and the Program of
Support for Leading Scientific Schools of Russia (NSh-1115.2003.2).

\vspace{1cm}
%%%%%%%%%%%%%%%%%%%%%%%%%%%%%%%%%%%%%%%%%%%%%%%%%
\begin{center}
 {\large  Appendix. The integration of the viscous
dissipation rate
           of oscillation energy over angular variables}
\end{center}
%%%%%%%%%%%%%%%%%%%%%%%%%%%%%%%%%%%%%%%%%%%%%%%%%%%%%%%%%%%%%%%%%
\vspace{0.5cm}

While calculating the angular integral in Eq.~(\ref{Eq24}), it is
convenient to introduce the notation
\begin{equation}
    \widetilde{\sigma}_{ik}\equiv \frac{\partial U_i}{\partial
    x_k}+\frac{\partial U_k}{\partial x_i} .
\end{equation}
Then, the part of Eq.~(\ref{Eq24}), associated with the shear
viscosity, can be written as
\begin{equation}
\label{EqA1}
    \frac {{\rm d}E}{{\rm d}t}=-\frac 14 \int_0^{R} \,\eta \,
     r^2\,{\rm d} r \int
    \left[\widetilde{\sigma}_{ik}\widetilde{\sigma}_{ki}^\ast
    -\frac 43 \left|\nabla{\bm U}\right|^2\right]\,{\rm d}\Omega
    =-\frac 14 \int_0^{R} r^2\, \eta
     \left(I_1-\frac 43 I_2\right)\,{\rm d} r,
\end{equation}
where $\,{\rm d} \Omega$ is a solid angle element,
\begin{equation}
    I_1\equiv
    \int    \widetilde{\sigma}_{ik}\widetilde{\sigma}_{ki}^\ast\,{\rm d}\,\Omega
    \quad\mbox {and}\quad  I_2\equiv
    \int \left|\nabla {\bm U}\right|^2\,{\rm d}\Omega.
\end{equation}
Here, we have assumed that the unperturbed star is spherically
symmetric, so that the shear viscosity does not depend on
angular variables. The integrals $I_1$ and $I_2$ have been taken
analytically for the
hydrodynamic velocities of the form ${\bm U}=\nabla\phi$, where
the velocity potential is $\phi={\rm e}^{\imath\omega
t}Y_{lm}(\theta,\varphi) F(r)$.

The integral $I_1$ can be taken if we write the components of the
tensor $\widetilde{\sigma}_{ik}$  in spherical coordinates (see, e.g.,
Ref.\ \cite{Gidrodinamika}). After this, the angular
integration is carried out analytically (using the properties of the
function $Y_{lm}(\theta,\varphi) $; see, e.g.,
Ref.\ \cite{Varshalovich}). This yields
\begin{eqnarray*}
I_1&=&4\,\left\{\left(F^{\prime\prime}\right)^2
    +2\,\frac{1+l(l+1)}{r^2}\,\left(F\,^\prime\right)^2
    -6\,\frac{l(l+1)}{r^3}\,F\,^\prime F
    +l(l+1)\,\frac{1+l(l+1)}{r^4}\,F^2\right\} \\
    &\approx&4\,\left\{\left(F^{\prime\prime}\right)^2
    +2\,\frac{l(l+1)}{R^2}\,\left(F\,^\prime\right)^2
    +l(l+1)\,\frac{l(l+1)}{R^4}\,F^2\right\},
\end{eqnarray*}
where the last equality is valid in the approximation
of plane-parallel layer.

Let us now consider the integral $I_2$. For this purpose, we write
the divergence of the velocity
\begin{eqnarray}
\nabla{\bm U}&=&\Delta\phi
    =\left(\frac 1{r^2}\,{\frac{\partial {}}{\partial {\,r}}}
        \,r^2\,{\frac{\partial {\,F}}{\partial {\,r}}}\, Y_{lm}(\theta,\varphi)
        +F\,\Delta_\Omega\, Y_{lm}(\theta,\varphi)\right)
    \,{\rm e}^{\imath \omega t}
    \nonumber \\
    &=&\left(F^{\prime\prime}+\frac{2\,F\,^\prime}{r}
        -\frac{l(l+1)}{r^2}\,F\right)\,Y_{lm}(\theta,\varphi)
    \,{\rm e}^{\imath \omega t}
    ,\label{EqA2}
\end{eqnarray}
where $\Delta_\Omega$ is the angular part of the Laplacian. The
integral $I_2$ is easily calculated,
\begin{equation}
I_2=\left(F^{\prime\prime}+\frac{2\,F\,^\prime}{r}
    -\frac{l(l+1)}{r^2}\,F\right)^2
    \approx
    \left(F^{\prime\prime}-\frac{l(l+1)}{R^2}\,F\right)^2,
\end{equation}
where the last equality is valid in the plane-parallel layer
approximation. As expected, Eq.~(\ref{EqA1}) does not depend
on the azimuthal number $m$ (due to the spherical symmetry of the
unperturbed star). It can be shown that ${\rm d}E/{\rm d}t \geq 0$ for all
allowed values $l = 0,\, 1,\, 2,\ldots$

%%%%%%%%%%%%%%%%%%%%%%%%%%%%%%%%%%%%%%%%%%%%%%%%%%%%%%%%%%%%%%%%%%%%%%%%%%%%%%%
                    
%
\end{document}